\date{Accepted XXX. Received YYY; in original form ZZZ}
\documentclass[fleqn,usenatbib]{mnras}
\usepackage{txfonts}

\usepackage[T1]{fontenc}
\usepackage{ae,aecompl, longtable, lscape}

\usepackage{color, soul, ulem,graphicx, amsbsy}
\usepackage{lipsum}
\newcommand{\msano}{{\rm M}_\odot~{\rm yr}^{-1}}
\newcommand{\e}[1]{\times 10^{#1}}
\newcommand{\ly}{Ly-$\alpha$}
\newcommand{\mdotsun}{\dot{M}_\odot}

\newcommand{\review}[1]{{\color{black} #1}}

\title{The dichotomy of atmospheric escape in AU Mic b}
\author[]{S.~Carolan$^1$, A.~A.~Vidotto$^1$, P.~Plavchan$^2$, C.~Villarreal D'Angelo$^{3,1}$, G.~Hazra$^1$
\\
$^1$School of Physics, Trinity College Dublin, the University of Dublin, Dublin-2, Ireland\\
$^2$Department of Physics and Astronomy, George Mason University, Fairfax, VA, 22030, USA \\
$^3$Observatorio Astron\'omico de C\'ordoba - Universidad Nacional de C\'ordoba. Laprida 854, X5000BGR. C\'ordoba, Argentina
}

\date{Accepted XXX. Received YYY; in original form ZZZ}

\pubyear{2020}

\begin{document}
\label{firstpage}
\pagerange{\pageref{firstpage}--\pageref{lastpage}}
\maketitle

%%%%%%%%%%%%%%%%%%%%%%%%%%%%%%%%%%%%%%%%%%%%%%%%%%%%%%%%%%%
\begin{abstract}
Here, we study the dichotomy of the escaping atmosphere of the newly discovered close-in exoplanet AU Mic b. On one hand, the high EUV stellar flux is expected to cause a strong  atmospheric escape in AU Mic b. On the other hand, the wind of this young star is believed to be very strong, which could reduce or even inhibit the planet's atmospheric escape. AU Mic is thought to have a wind mass-loss rate that is up to $1000$ times larger than the solar wind mass-loss rate ($\mdotsun$). To investigate this dichotomy, we perform 3D hydrodynamics simulations of the stellar wind--planetary atmosphere interactions in the AU Mic system and predict the synthetic \ly\ transits of AU Mic b. We systematically vary the stellar wind mass-loss rate from a `no wind' scenario to up to a stellar wind with a mass-loss rate of $1000~\mdotsun$. We find that, as the stellar wind becomes stronger, the planetary evaporation rate decreases from $6.5\e{10}$ g/s to half this value. With a stronger stellar wind, the atmosphere is forced to occupy a smaller volume, affecting transit signatures. Our predicted \ly\ absorption drops from $\sim 20\%$, in the case of `no wind' to barely any \ly\ absorption in the extreme stellar wind scenario. Future \ly\ transits could therefore place constraints not only on the evaporation rate of AU Mic b, but also on the mass-loss rate of its host star. 

\end{abstract}
%%%%%%%%%%%%%%%%%%%%%%%%%%%%%%%%%%%%%%%%%%%%%%%%%%%%%%%%%%%
\begin{keywords}
stars: planetary systems: AU Mic --  planet-star interactions %stars: winds, outflows -- 
\end{keywords}

\vspace{-0.5cm}
%%%%%%%%%%%%%%%%%%%%%%%%%%%%%%%%%%%%%%%%%%%%%%%%%%%%%%%%%%%%
\section{Introduction}\label{sec.intro}
AU Microscopii (AU Mic) is the second closest pre-main sequence star to the solar system (9.79 pc). With an age of approximately 22~Myr, it is orbited by an edge-on debris disk, within which lies the recently discovered warm Neptune AU Mic b \citep{Plavchan2020}. Detections of such young exoplanets are still rare, given that young stars like AU Mic pose many observational challenges for planet detection, such as the presence of spots and frequent flares. AU Mic b can therefore provide unique insights into newly formed planets and their atmospheres. 

Due to its youth and activity, AU Mic emits a large flux of high-energy photons in the extreme-ultraviolet (EUV). Combined with the small orbital distance (0.066~au), the estimated EUV flux impinging on AU Mic b is $4.7\e{3}$~erg\,cm$^{-2}$\,s$^{-1}$, and can be as high as $2.2\e{4}$~erg\,cm$^{-2}$\,s$^{-1}$ when the star is in flaring state (stellar fluxes from \citealt{Chadney2015}). These values are 10--50 times larger than the estimated flux received in HD209458b of 450~erg\,cm$^{-2}$\,s$^{-1}$ \citep{MurrayClay2009}, a hot Jupiter that shows strong atmospheric evaporation \citep{VM2003}. By analogy, one would expect that AU Mic b would be strongly evaporating. Additionally, due to the youth of the system, the planet likely still has its primordial atmosphere, which would be mainly composed of hydrogen. A strong evaporation of a hydrogen-rich atmosphere, such as the one AU Mic b could host, is better probed in hydrogen lines, such as in \ly\ or the Balmer series, through spectroscopic transits.

As the high-energy flux is deposited in the \review{thermosphere}, the atmosphere is heated and expands. As a consequence, the atmosphere escapes the planet in the form a photo-evaporative outflow \citep[e.g.][]{MurrayClay2009}. On its journey up, the evaporating atmosphere is accelerated from a subsonic to a supersonic flow that eventually crosses the Roche lobe and escapes from the planet.

One important point to consider when studying planetary evaporation is that the escaping atmospheres do not expand into an empty space, but rather the atmosphere pushes its way into the stellar wind. The stellar wind of cool dwarfs consists of a hot, ionised plasma, that is embedded in the stellar magnetic field \citep{Vidotto2015}. Stellar winds can affect atmospheric evaporation of close-in exoplanets \citep[e.g.,][]{McCann2019, Shaikhislamov2020, Carolan2020}. In particular, the stronger the stellar wind is, the larger is the pressure it exerts in the planetary atmosphere. In a simplified way, we can think of this as the interaction of two fluids. The point where the two fluids meet is determined by pressure balance. Therefore, the stronger the stellar wind is, the point where balance is achieved is reached deeper in the evaporating atmosphere, which, as a consequence, is forced to occupy a \mbox{smaller volume}.

If the interaction happens so deep in the planetary atmosphere, where the planetary  outflow is still subsonic, the stellar wind could substantially reduce the evaporation  \citep{Christie2016, Vidotto2020}. Using 3D hydrodynamics simulation, \citet{Carolan2020} performed a systematic study of the effects of the stellar wind on the evaporation rate of a typical hot Jupiter. They showed that for weaker stellar winds, the reduction in planetary escape rate was very small. Nevertheless, because the atmosphere was forced to occupy a smaller volume, spectroscopic transit signatures were substantially affected. The atmospheric escape rates remained approximately constant ($\simeq 5.5\times 10^{11}$g/s), while its \ly\ transit absorption changed from 24\% to 14\% as the stellar wind mass-loss rate was only moderately increased from `no wind' to a wind with a mass-loss rate that is 10 times the solar value of $\dot{M}_\odot=2\e{-14}~\msano$. However, as the stellar wind became stronger than that, a more substantial reduction in evaporation rates was seen, in particular  after this interaction started to occur below the sonic surface of the planetary outflow. For a stellar wind mass-loss rate of $100~\dot{M}_\odot$, the evaporation rate had reduced 65\% and the absorption in the \ly\ line went down to less than 5\%.

This leads to a dichotomy for the AU Mic system.  While, due to the large EUV flux impinged on the atmosphere, the evaporation rate of a close-in planet is expected to be very strong during its youth, the stellar wind is also stronger at young ages \citep{VidottoDonati2017, Carolan2019}. AU Mic in particular is thought to have a wind mass-loss rate that is larger than solar. Theoretical estimates range from $10~\dot{M}_\odot$ \citep{2009ApJ...698.1068P} to $1000~\dot{M}_\odot$ \citep{2006ApJ...648..652S, 2017ApJ...848....4C}.

To investigate what could be possibly happening in the AU Mic system and to guide whether strong evaporation could be detected in \ly\ transits, we study  how the wind of AU Mic could affect the evaporation rate of AU Mic b and its predicted transit. For that, we use 3D hydrodynamics simulations followed by synthetic line profile calculations \review{that} investigate the effect increasing the strength of the stellar wind has on \ly\ transits of AU Mic b. 

\vspace{-0.5cm}
\section{Atmospheric Escape and Synthetic \ly\ transit Models}\label{sec.model}
We use the model presented in \citet{Carolan2020} to study the effects of the stellar winds on AU Mic b. Here, we briefly introduce the model and point the reader to \citet{Carolan2020} for further details. Our model uses  the Space Weather Modelling Framework \citep{Toth-swmf} to perform the 3D hydrodynamics simulation of the interaction betwen the stellar wind and the planetary atmosphere. The planet is centred in the 3D grid and the stellar wind is injected through an outer boundary. We assume these two flows are isothermal, with the stellar wind having a temperature of 2~MK and the planetary outflow a temperature of $5300$K (our choice of temperature is discussed below). We use a rectangular grid that extends from $[-50, 50]~r_p$ in the $x$ and $y$ directions and  $[-32, 32]~r_p$ in the $z$ direction, where $r_p$ is the radius of the planet. The orbital plane is in the $xy$ plane and the orbital spin axis is along positive $z$. AU Mic b is in a prograde orbit nearly perpendicular to the stellar spin axis \citep{Martioli2020}. Our grid contains $\sim$16 million cells and has a non-uniform resolution, with the highest resolution of $1/16~r_p$ within a radius of $5~r_p$, and gradually coarser towards the edge of the grid.  We solve for the mass density $\rho$, velocity $\vec{u} =[u_x, u_y, u_z]$ and thermal pressure $P$ in the  frame corotating with the planet, which is assumed to have the same rotational angular velocity as the orbital angular velocity, for simplicity. We solve a set of coupled hydrodynamic equations. The momentum  equation is 
\begin{equation}\label{eq.mom}
    \frac{\partial(\rho\vec{u})}{\partial t} + \nabla \cdot [\rho \vec{u} \vec{u} + PI] = \rho \vec{g} -\frac{\rho GM_{*}}{(r-a)^2} \hat{R} - \rho\vec{\Omega} \times (\vec{\Omega}\times\vec{R})-2\rho(\vec{\Omega} \times \vec{u}) , 
\end{equation}
where $I$ is the identity matrix, $\vec{g}$ the acceleration due to the planet's gravity, $G$  the gravitational constant, and $M_*$ is the mass of the star. $\vec{r}$ is the position vector relative to the planet, $\vec{a}$  the position of the star relative to the planet, $\vec{\Omega}$  the orbital rotation rate, and $\vec{R}$ is the position vector relative to the star. In the right-hand side of Eq.~(\ref{eq.mom}) we have the planetary gravitational force,  stellar gravity, and the centrifugal  and  Coriolis forces. The energy conservation equation is
\begin{equation}
    \frac{\partial \epsilon}{\partial t} + \nabla \cdot [\vec{u}(\epsilon + P )] = \rho \bigg( \vec{g} -\frac{GM_{*}}{(r-a)^2} \hat{R} - \vec{\Omega} \times (\vec{\Omega}\times\vec{R}) \bigg) \cdot \vec{u},
\end{equation}
where $\epsilon = {\rho u^2}/{2} + {P}/({\gamma -1})$. We take $\gamma=1.001$, which implies that the flows are nearly isothermal. We assume an ideal gas, where the thermal pressure is $P = \rho k_B T/(\mu m_p)$, where $k_B$ is the Boltzmann constant, and  $\mu$ is the  mean mass per particle and $m_p$ is the mass of the proton. Our 3D simulations assume fully ionised hydrogen flows, thus $\mu=0.5$.  The set of equations is closed with the mass conservation equation 
\begin{equation}
    \frac{\partial \rho}{\partial t} + \nabla \cdot (\rho \vec{u}) = 0.
\end{equation}

\begin{figure}
    \centering
    \includegraphics[width=0.9\columnwidth]{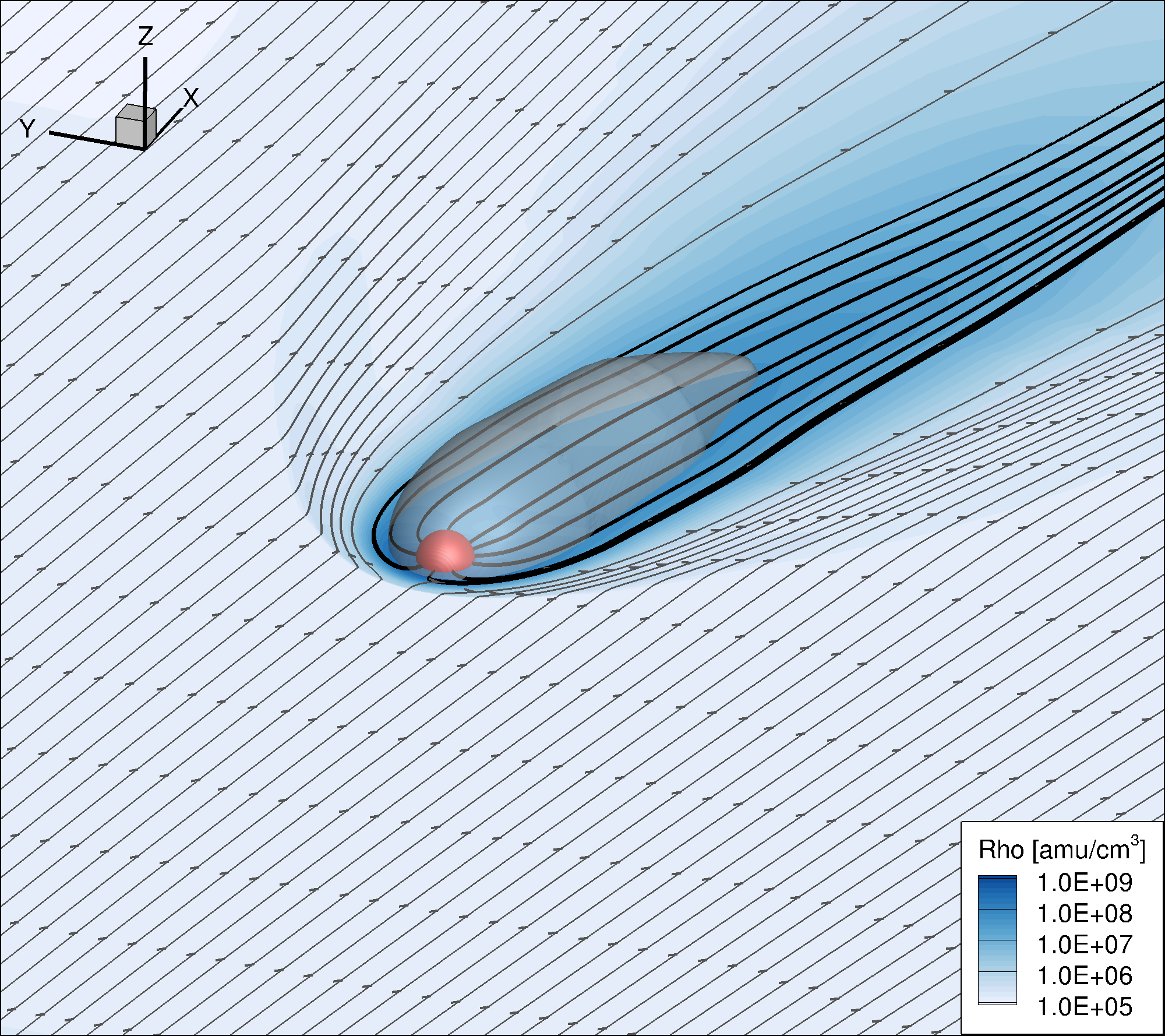}
    \caption{Atmospheric escape of AU Mic b, when it interacts with a stellar wind with $\dot{M} = 100\dot{M}_\odot$. The stellar wind is injected in the negative $x$. Its streamlines are shown in grey, while the black streamlines represent the velocity field of the planetary outflowing atmosphere. The density is shown in the equatorial plane and the grey surface around the planet shows the region used in the synthetic transits (with temperature $\lesssim 1.1 T_p$, Section \ref{sec.model}).}
    \label{fig:3dmodel}
\end{figure}

We assume a planetary mass of $0.69~M_{\rm Nep}$ and radius of $1.08~R_{\rm Nep}$. The stellar mass and radius are $M_* = 0.5~M_{\odot}$ and $R_* = 0.75~R_{\odot}$. The orbital distance is $a = 0.066$~au, transit duration is 3.5h and the impact parameter is $0.16R_\star$. All these values are from \citet{Plavchan2020}, with the exception of the planetary mass that is from Plavchan et al. (in prep.), \review{and was obtained from additional radial velocity measurements}. Figure \ref{fig:3dmodel} shows the output of one of our simulations, after reaching steady state.

To calculate the \ly\ transit profiles, we use the ray tracing technique detailed in \citet{Allan2019}. Stellar rays are shot through the planetary material, which is represented by the volume entailed by the grey surface in Figure \ref{fig:3dmodel}. We  calculate the velocity-dependent optical depth $\tau_v$ of this material along the line-of-sight and integrate over  all rays that are transmitted through the atmosphere to obtain the velocity-dependent transit depth
\begin{equation}\label{eq.deltaF}
    \Delta F_v = {\int\int (1-e^{-\tau_v})\,  dydz}/({\pi R_*^2}).
\end{equation}
The optical depth of  the \ly\ transit requires the density of neutrals $n_n$, such that $\tau_v =  \int n_n \sigma_v \phi_v\, dx$, where $\phi_v$ is the Voigt line profile, and $\sigma= 0.01103$~cm$^{2}$~Hz is the \ly\ absorption cross section at line centre. 
Because our 3D model  does not treat the neutral material of the planetary outflow, we use a post-processing technique to estimate its ionisation fraction $f_i$. This is done using the 1D atmospheric escape model of \citet{Allan2019}, where we assume an EUV luminosity of $1.5\e{-5}~L_\odot$ appropriate for the quiescent state of AU Mic \citep{Chadney2015}. With this, we derive the density of neutrals as $n_n = n_p (1-f_i)/f_i$, where $n_p$ is the proton density from our 3D simulations. 

The results of our 1D model are also used to constrain the free parameters in the 3D simulations, namely the atmospheric base temperature and density. The 1D model solves the energy equation of the planetary outflow assuming photoionisation by stellar EUV photons and \ly\ cooling \citep{Allan2019}. As a result, the atmospheric temperature varies from 1000K at $1r_p$ to nearly 8000~K at $\sim 2r_p$, and cools beyond that. We pick an intermediate temperature of $5300$~K for our 3D model. We also chose the base density of our 3D simulations such that it matches the predicted escape rate of $6.5\e{10}$~g/s from the 1D model. 

\vspace{-0.5cm}
\section{Results: systematic variation of the stellar wind strength}\label{sec.results}
We perform 5 simulations where we systematically vary the stellar wind mass-loss rate: $\dot{M}=0$ (no wind), 1, 10, 100 and $1000~\dot{M}_\odot$. We use a temperature of $2$MK for the stellar wind, so that the stellar wind \review{is always injected with the same velocity (about 540 km/s).} As this is a thermally-driven wind, the wind velocity is independent of the density, so changing $\dot{M}$ while the velocity structure is constant solely changes the density profile of the injected stellar wind. 

\begin{figure*}
    \centering
    \includegraphics[width=0.9\textwidth]{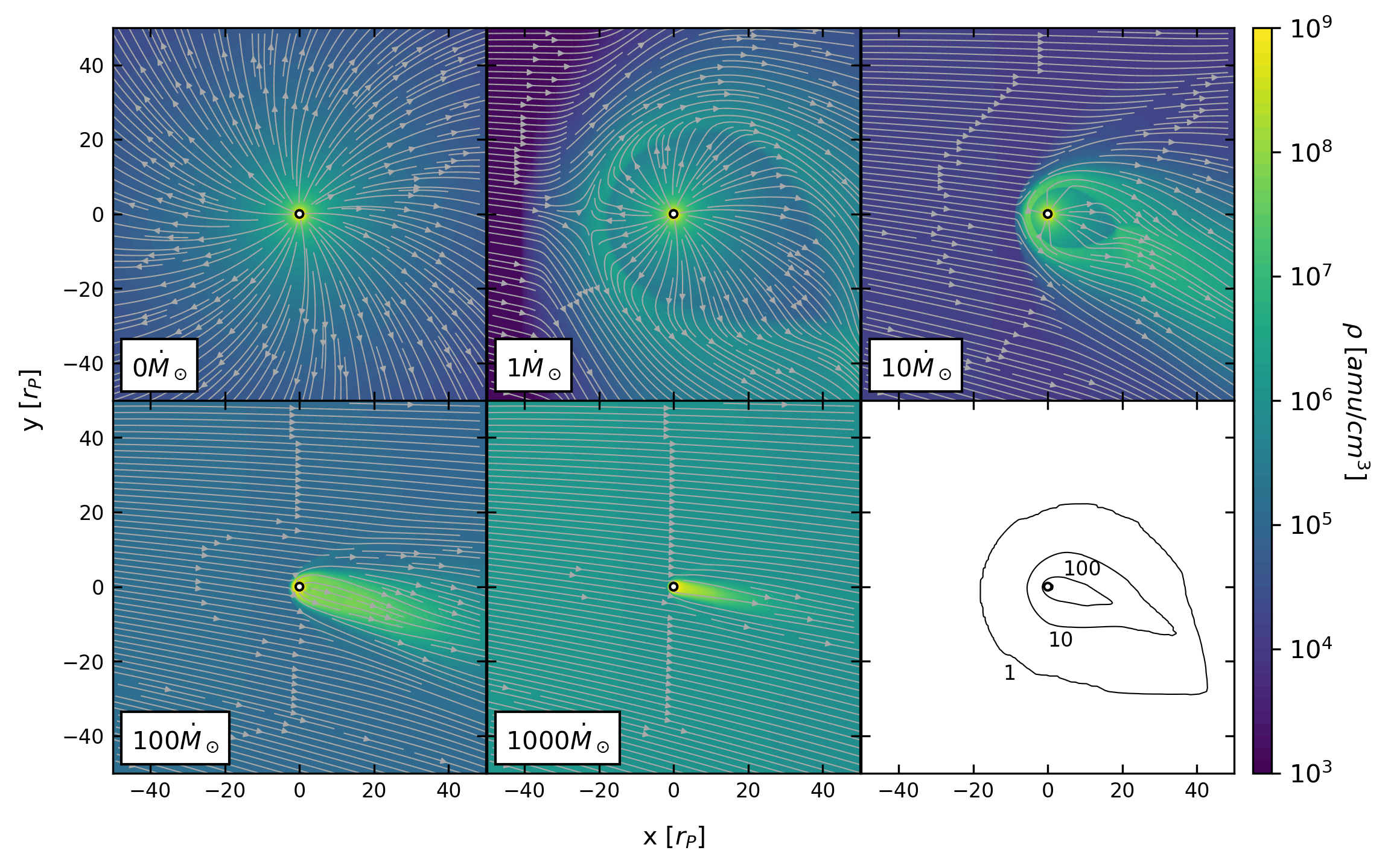}
    \caption{Density structure and velocity streamlines in the planet's reference frame for the 5 simulations we run for varying stellar wind $\dot{M}$, quoted on the first five panels. The planet is shown at the center of the grid on the orbital plane.  As the stellar wind that is injected in the negative $x$ boundary has a larger $\dot{M}$, the escaping atmosphere of AU Mic b is forced into smaller volumes. The last panel shows the iso-contours of temperature at approximately the temperature we adopt for the planetary atmosphere. Material within these contours belong to the planet and are used in the synthetic calculations of the \ly\ transit profiles. The numbers shown next to each iso-contour represent the stellar wind mass-loss rate in $\mdotsun$. The inner-most contour is for the case with 1000$\dot{M}_\odot$ (label not shown).}
    \label{fig:models}
\end{figure*}

Figure \ref{fig:models} shows the orbital slice of each of these simulations. We see that as the stellar wind mass-loss rate (and thus its ram pressure) is increased, the escaping atmosphere is confined closer to the planet, and forced to occupy a smaller volume. The position where the interaction happens eventually disrupts the sonic surface (originally at $1.3 r_p$) of the escaping atmosphere, such that parts of the planetary outflow, especially in the dayside, can no longer accelerate to supersonic speeds. When this happens, we see a stronger decrease in the escape rate of the planetary atmosphere. 

We calculate the escape rate by integrating the mass flux through concentric spheres (with areas $A$) around the planet: $\dot{m} = \oint_A  \rho \vec{u} \cdot dA$.
These values are given in Table \ref{tab:models}, where we see that the escape rate is unaffected in the 1 and 10 $\dot{M}_\odot$ models, and they are very similar to the values we obtain in the `no wind' model. In the 100 $\dot{M}_\odot$ model the escape rate has decreased slightly from $6.5\times10^{10}$ to $5.9\times10^{10}$ g/s. This is the first of our computed models where the wind is capable of sufficiently confining the escaping atmosphere such that the dayside sonic surface is disrupted. The dayside flow is no longer able to reach supersonic speeds. Material continues to outflow from the planet but they are funnelled back towards the comet-like tail (better seen in Figure \ref{fig:3dmodel}). The nightside sonic surface remains unaffected, so only a small decrease in escape rate is found in this model. This is not the case in the 1000 $\dot{M}_\odot$ model, where the stellar wind confines the escaping atmosphere such that the sonic surface on all sides of the planet is affected. This results in a $50\%$ lower escape rate, when compared to other models.

\begin{table}
    \centering
        \caption{Simulation results showing  the stellar wind mass-loss rate ($\dot{M}$), planetary atmosphere escape rate ($\dot{m}$), absorption in the blue ([-100,-36] km/s) and red ([36,100] km/s) wings of the \ly\ line and the sum of these absorptions.}
    \label{tab:models}
    \begin{tabular}{lcccc}
        \hline
        $\dot{M}~(\dot{M}_\odot)$ & $\dot{m}$ ($10^{10}$g/s) & $\Delta F_{\rm blue}$ (\%) & $\Delta F_{\rm red}$ (\%)  &$\Delta F_{\rm tot}$ (\%) \\
         \hline
0 (no-wind)   &6.5&9.9 &10  &20\\
1   &6.5&9.8 &5.8 &16\\
10  &6.3&4.1 &1.4 &5.5\\
100 &5.9&1.4 &1.1 &2.5\\
1000&3.2&0.25&0.25&0.5\\
        \hline
    \end{tabular}
\end{table}

We investigate the effect this confinement has on the \ly\ transit by performing synthetic transit calculations. To ensure that the material we use in the ray tracing computation is planetary, we use a temperature cut-off that is 10\% higher than the planetary outflow and follow the description presented in Section \ref{sec.model}.  The contours of this temperature cut-off is seen in the last panel of Figure \ref{fig:models} for the orbital plane, further illustrating how the confinement of the planetary atmosphere varies in each model.

\begin{figure*}
    \centering
    \includegraphics[height=0.2\textheight]{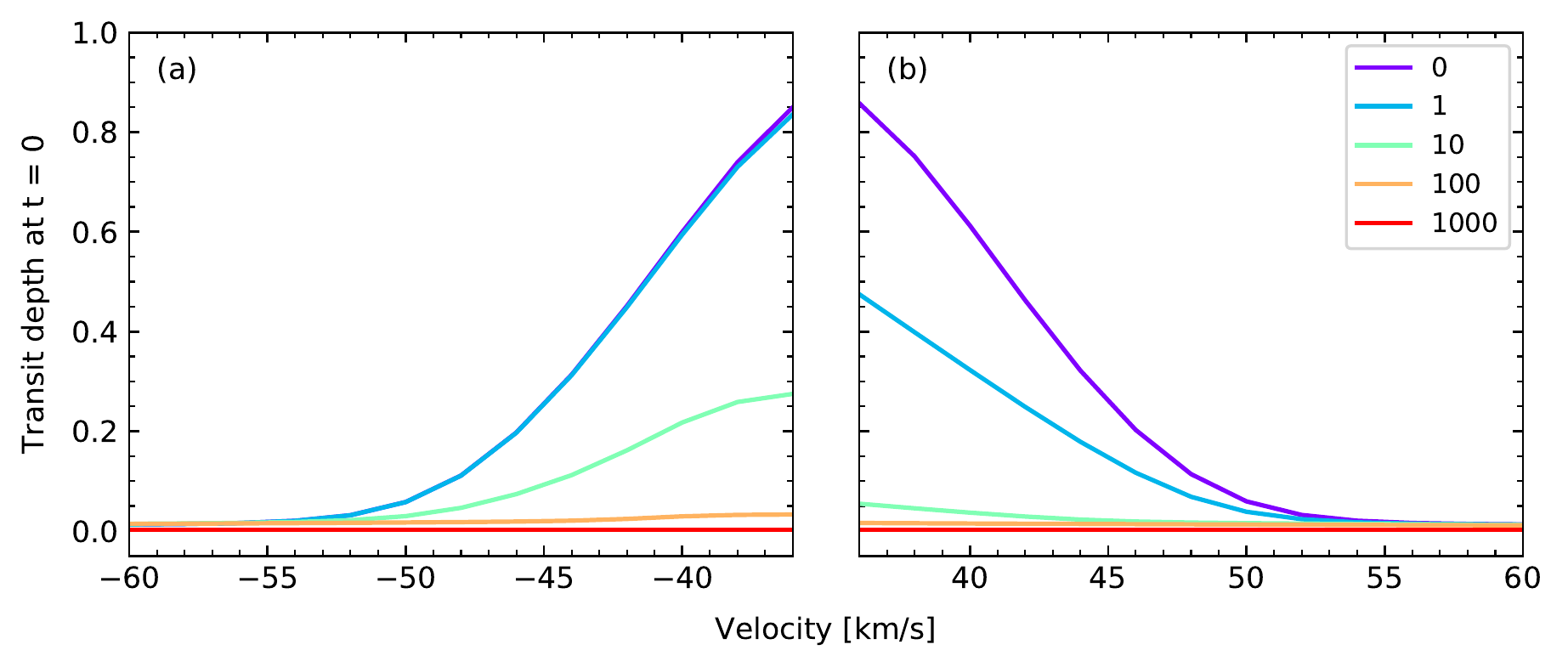}     \includegraphics[height=0.2\textheight]{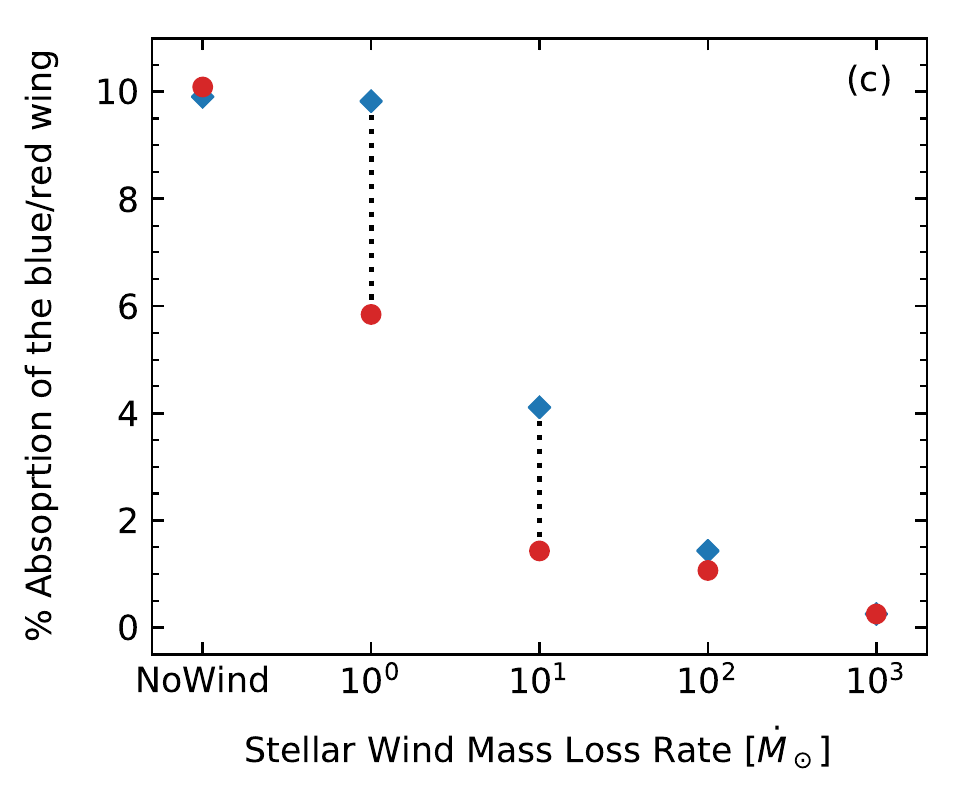}
    \caption{(a) and (b) Transit depth of the \ly\ line computed at mid-transit for the blue  ($\leq - 36$ km/s) and red  ($\geq36$ km/s) wings, respectively, as a function of Doppler velocity. (c) Integrated absorption in the blue (blue diamonds) and red (red circles) wings of the \ly\ line calculated at mid-transit, as a function of the stellar wind mass-loss rate. Table \ref{tab:models} shows these values and the total absorption.}
    \label{fig:raytrace}
\end{figure*}

In none of our models the line centre is saturated (i.e., 100\% absorption), but models with $0$ and $1~\mdotsun$ reach more than 95\% absorption at line centre. However, given the line centre of the \ly\ line is contaminated by geocoronal emission and interstellar absorption \review{(assuming the stellar and ISM radial velocities are $\simeq 0$)}, we do not consider the line centre [-36, 36] km/s in our results presented next. The blue [-100, -36] km/s and red [36, 100] km/s wings of the \ly\ line are shown in Figure \ref{fig:raytrace}a and b, respectively. The absorption computed in these velocity intervals $\int_{v_i}^{v_f} \Delta F_v dv/(v_f-v_i)$ are shown in Table \ref{tab:models} and Figure \ref{fig:raytrace}c, where $v_i$ and $v_f$ are initial and final velocities in the ranges quoted above.

The no-wind model is the only case where the line profile is nearly symmetric in both wings. Line asymmetry is already seen in model $1\mdotsun$. Despite the escape rate of the planet remaining unchanged for models $0$, $1$ and $10\mdotsun$, the \ly\ absorption  has changed significantly when compared to the no-wind model. For the $10 \dot{M}_\odot$ model, we see a greater reduction in the red wing absorption, as the planetary flow towards the star is suppressed by the stellar wind. Though we still see blue wing absorption in the $10 \dot{M}_\odot$ model, it has been significantly reduced compared to the 0 and $1\mdotsun$ models. This is because the stronger stellar wind reduces the volume of the comet-like tail, which contains most of the blue shifted absorbing material. The volume of absorbing material is further reduced in the $100 \dot{M}_\odot$ model such that very little blue wing absorption is found, and essentially no detectable red wing absorption. The $1000 \dot{M}_\odot$ stellar wind confinement has not only reduced the escape rate of the planet by 50\% but also completely masked the observational signatures of this escape in \ly , as we find no red or \mbox{blue wings absorption}.
 
\vspace{-0.5cm}
\section{Discussion and Conclusion}\label{sec.conclusions}
We investigated here the dichotomy of atmospheric escape in the newly discovered exoplanet AU Mic b. On one hand, the high EUV  flux of young host stars is expected to cause strong atmospheric escape \review{\citep{Kubyshkina2018}}. On the other hand, the star is expected to have a strong stellar wind  \citep[10 to $10^3 \mdotsun$, ][]{2006ApJ...648..652S, 2009ApJ...698.1068P}, which can reduce evaporation in the planet \citep{Vidotto2020, Carolan2020}. To investigate this dichotomy, we modeled the interaction between the  wind of AU Mic with the escaping atmosphere of AU Mic b, by performing 3D hydrodynamics simulations of the system. We considered a number of stellar wind mass-loss rates from $\dot{M} = 0$ to $10^3\mdotsun$. We found that increasing $\dot{M}$ confines the escaping planetary atmosphere, which occupies a smaller volume. When this confinement disrupts the sonic surface of the planetary outflow, we see a more substantial reduction in the escape rate. For the models with 0, 1 and 10 $\dot{M}_\odot$ the escape rate is unaffected and remains $6.5\times10^{10}$ g/s. In the 100 $\dot{M}_\odot$ model, the escape rate  decreases slightly from $6.5\times10^{10}$ to $5.9\times10^{10}$ g/s. However, for the 1000 $\dot{M}_\odot$ model, escape rates is reduced by $50\%$, with a value of $3.2\times10^{10}$ g/s.
\review{From an evolutionary point of view, a factor of 2 in the escape rate is negligible. The reduction is likely to be more important at young ages, when the stellar wind is stronger, but it is worth recalling that the evaporation of the planet is also stronger at younger ages, so investigating which process ``wins'' (wind vs EUV flux) is an important point to consider in future studies of planetary evolution.}

The reduction in evaporation affects differently  \ly\ transit absorption. We calculated synthetic \ly\  line profiles at mid-transit and found that, although we still see blue wing absorption in the $10 \mdotsun$ model, it is significantly smaller than the 0 and $1\mdotsun$ models. This happens even though no appreciable reduction is seen in the escape rate of these 3 models. This is because the stronger stellar wind reduces the volume of the comet-like tail, which contains most of the blueshifted absorbing material. For the $10^3\dot{M}_\odot$ model, we found almost no \ly\ absorption, as most of the absorbing material is confined to a small volume around the planet. 
\review{Our models do not consider charge-exchange process, which converts a low velocity planetary neutral atom into an ion and, a high velocity stellar wind ion into an energetic neutral atom \citep{Shaikhislamov2020}. The net effect of this process is to shift absorption in the \ly\ centre to the blue wing. Given the already low absorption around line centre in the $>100\mdotsun$ models (low volume occupied by absorbing material) shifting this to higher velocities by charge-exchange will not greatly alter absorption at line wings. Charge-exchange can be more important for models with lower $\dot{M}$, given the volume containing neutrals is much larger \mbox{than the higher $\dot{M}$ models.}}

A few years ago, \citet{Chadney2015} predicted escape rates of a fictitious planet orbiting AU Mic. They assumed a hot-Jupiter planet similar to HD209458b at 0.2au, and obtained a strong evaporation rate of $1.2\e{10}$~g/s. Due to  differences in the planet parameters, a comparison between our results and theirs is not straightforward. If we were to `move' their fictitious planet to the orbit of AU Mic b, we estimate a factor of $(0.2/0.066)^2\simeq 9$ (i.e., linear with EUV flux) increase in their escape rates, bringing their estimates to $10^{11}$ g/s. This evaporation rate is about a factor of 2 larger than our `no wind' model, but is on the same order of magnitude. Note though that we are comparing two different planets here -- a fictitious hot Jupiter and AU Mic b, which is a warm Neptune, and that the differences in planetary gravity will affect escape rates \citep{Allan2019}. 

The numbers we quoted in this paper should be used with care, as they are dependent on assumptions we made for the system, such as the planetary escape rate in the no-wind model, which we set from an assumed EUV flux, and the stellar wind temperature, which we assumed is a typical coronal-like temperature of 2 MK. 
\review{For example, the evaporation rate of a planet that has a sonic surface at a larger distance is more easily affected by the stellar wind. Likewise, a stellar wind that has a larger ram pressure more easily disrupts the planet's sonic surface, affecting  more the evaporation. \citet{Carolan2020} further discuss how different system characteristics affect the reduction of escape rate and absorption.} Another assumption we made is that the stellar wind is spherically symmetric (purely radial velocity and isotropic mass flux). If the star has a complex magnetic field topology, the stellar wind will not be isotropic, as the field geometry leads to a non-homogeneous stellar wind  along the planetary orbital path \citep{Vidotto2015}. Even if we were to adopt slightly different values for the planet escape rate or stellar wind properties, the general conclusions we found here should remain valid. Namely, we concluded that should future \ly\ observations detect solely blue wing absorption during the transit of AU Mic b (i.e., little or no redshifted absorption), the stellar wind mass-loss rate of AU Mic can be estimated to be $\sim 10\dot{M}_\odot$. A redshifted absorption would imply mass-loss rates $\lesssim 10\mdotsun$. Should future observations find a non-detection in \ly\ transits, we propose that this could be due to stellar wind confinement of the escaping atmosphere. In this case, our models would allow us to place a lower limit on the mass-loss rate of AU Mic of $\gtrsim 100\dot{M}_\odot$. This would help clarify whether AU Mic has a moderately strong wind ($10\mdotsun$) or substantially stronger ($10^3\mdotsun$). 

\smallskip

\noindent \textbf{Acknowledgements:} We acknowledge funding from the ERC grant 817540, ASTROFLOW and computational facilities and support from SFI/HEA ICHEC. This work used BATS-R-US developed at the U.~of Michigan CSEM and made available through the NASA CCMC. \review{We thank the referee, Dr Fossati, for constructive comments.}
\smallskip

\noindent \textbf{Data Availability:} The data described in this article will be shared on reasonable request to the corresponding author.

%@arxiver{3d_fig_T_surface.png,six_panel.png,blue_and_red.pdf}

%%%%%%%%%%%%%%%%%%%%%%%%%%%%%%%%%%%%%%%%%%%%%%%%%%


\vspace{-0.5cm}
\label{lastpage}
\bibliography{mybib}

\begin{thebibliography}{}
\makeatletter
\relax
\def\mn@urlcharsother{\let\do\@makeother \do\$\do\&\do\#\do\^\do\_\do\%\do\~}
\def\mn@doi{\begingroup\mn@urlcharsother \@ifnextchar [ {\mn@doi@}
  {\mn@doi@[]}}
\def\mn@doi@[#1]#2{\def\@tempa{#1}\ifx\@tempa\@empty \href
  {http://dx.doi.org/#2} {doi:#2}\else \href {http://dx.doi.org/#2} {#1}\fi
  \endgroup}
\def\mn@eprint#1#2{\mn@eprint@#1:#2::\@nil}
\def\mn@eprint@arXiv#1{\href {http://arxiv.org/abs/#1} {{\tt arXiv:#1}}}
\def\mn@eprint@dblp#1{\href {http://dblp.uni-trier.de/rec/bibtex/#1.xml}
  {dblp:#1}}
\def\mn@eprint@#1:#2:#3:#4\@nil{\def\@tempa {#1}\def\@tempb {#2}\def\@tempc
  {#3}\ifx \@tempc \@empty \let \@tempc \@tempb \let \@tempb \@tempa \fi \ifx
  \@tempb \@empty \def\@tempb {arXiv}\fi \@ifundefined
  {mn@eprint@\@tempb}{\@tempb:\@tempc}{\expandafter \expandafter \csname
  mn@eprint@\@tempb\endcsname \expandafter{\@tempc}}}

\bibitem[\protect\citeauthoryear{{Allan} \& {Vidotto}}{{Allan} \&
  {Vidotto}}{2019}]{Allan2019}
{Allan} A.,  {Vidotto} A.~A.,  2019, \mn@doi [\mnras] {10.1093/mnras/stz2842},
  \href {https://ui.adsabs.harvard.edu/abs/2019MNRAS.490.3760A} {490, 3760}

\bibitem[\protect\citeauthoryear{{Carolan}, {Vidotto}, {Loesch}  \&
  {Coogan}}{{Carolan} et~al.}{2019}]{Carolan2019}
{Carolan} S.,  {Vidotto} A.~A.,  {Loesch} C.,   {Coogan} P.,  2019, \mn@doi
  [\mnras] {10.1093/mnras/stz2422}, \href
  {https://ui.adsabs.harvard.edu/abs/2019MNRAS.489.5784C} {489, 5784}

\bibitem[\protect\citeauthoryear{{Carolan}, {Vidotto}, {Villarreal D'Angelo}
  \& {Hazra}}{{Carolan} et~al.}{2020}]{Carolan2020}
{Carolan} S.,  {Vidotto} A.~A.,  {Villarreal D'Angelo} C.,   {Hazra} G.,  2020,
  \mnras, submitted

\bibitem[\protect\citeauthoryear{{Chadney}, {Galand}, {Unruh}, {Koskinen}  \&
  {Sanz-Forcada}}{{Chadney} et~al.}{2015}]{Chadney2015}
{Chadney} J.~M.,  {Galand} M.,  {Unruh} Y.~C.,  {Koskinen} T.~T.,
  {Sanz-Forcada} J.,  2015, \mn@doi [\icarus] {10.1016/j.icarus.2014.12.012},
  \href {https://ui.adsabs.harvard.edu/abs/2015Icar..250..357C} {250, 357}

\bibitem[\protect\citeauthoryear{{Chiang} \& {Fung}}{{Chiang} \&
  {Fung}}{2017}]{2017ApJ...848....4C}
{Chiang} E.,  {Fung} J.,  2017, \mn@doi [\apj] {10.3847/1538-4357/aa89e6},
  \href {https://ui.adsabs.harvard.edu/abs/2017ApJ...848....4C} {848, 4}

\bibitem[\protect\citeauthoryear{{Christie}, {Arras}  \& {Li}}{{Christie}
  et~al.}{2016}]{Christie2016}
{Christie} D.,  {Arras} P.,   {Li} Z.-Y.,  2016, \mn@doi [\apj]
  {10.3847/0004-637X/820/1/3}, \href
  {https://ui.adsabs.harvard.edu/abs/2016ApJ...820....3C} {820, 3}

\bibitem[\protect\citeauthoryear{{Kubyshkina}, {Lendl}, {Fossati}, {Cubillos},
  {Lammer}, {Erkaev}  \& {Johnstone}}{{Kubyshkina}
  et~al.}{2018}]{Kubyshkina2018}
{Kubyshkina} D.,  {Lendl} M.,  {Fossati} L.,  {Cubillos} P.~E.,  {Lammer} H.,
  {Erkaev} N.~V.,   {Johnstone} C.~P.,  2018, \mn@doi [\aap]
  {10.1051/0004-6361/201731816}, \href
  {https://ui.adsabs.harvard.edu/abs/2018A&A...612A..25K} {612, A25}

\bibitem[\protect\citeauthoryear{{Martioli} et~al.,}{{Martioli}
  et~al.}{2020}]{Martioli2020}
{Martioli} E.,  et~al., 2020, arXiv e-prints, \href
  {https://ui.adsabs.harvard.edu/abs/2020arXiv200613269M} {p. arXiv:2006.13269}

\bibitem[\protect\citeauthoryear{{McCann}, {Murray-Clay}, {Kratter}  \&
  {Krumholz}}{{McCann} et~al.}{2019}]{McCann2019}
{McCann} J.,  {Murray-Clay} R.~A.,  {Kratter} K.,   {Krumholz} M.~R.,  2019,
  \mn@doi [\apj] {10.3847/1538-4357/ab05b8}, \href
  {https://ui.adsabs.harvard.edu/abs/2019ApJ...873...89M} {873, 89}

\bibitem[\protect\citeauthoryear{{Murray-Clay}, {Chiang}  \&
  {Murray}}{{Murray-Clay} et~al.}{2009}]{MurrayClay2009}
{Murray-Clay} R.~A.,  {Chiang} E.~I.,   {Murray} N.,  2009, \mn@doi [\apj]
  {10.1088/0004-637X/693/1/23}, \href
  {https://ui.adsabs.harvard.edu/abs/2009ApJ...693...23M} {693, 23}

\bibitem[\protect\citeauthoryear{{Plavchan}, {Werner}, {Chen}, {Stapelfeldt},
  {Su}, {Stauffer}  \& {Song}}{{Plavchan} et~al.}{2009}]{2009ApJ...698.1068P}
{Plavchan} P.,  {Werner} M.~W.,  {Chen} C.~H.,  {Stapelfeldt} K.~R.,  {Su}
  K.~Y.~L.,  {Stauffer} J.~R.,   {Song} I.,  2009, \mn@doi [\apj]
  {10.1088/0004-637X/698/2/1068}, \href
  {https://ui.adsabs.harvard.edu/abs/2009ApJ...698.1068P} {698, 1068}

\bibitem[\protect\citeauthoryear{{Plavchan} et~al.,}{{Plavchan}
  et~al.}{2020}]{Plavchan2020}
{Plavchan} P.,  et~al., 2020, \mn@doi [\nat] {10.1038/s41586-020-2400-z}, \href
  {https://ui.adsabs.harvard.edu/abs/2020Natur.582..497P} {582, 497}

\bibitem[\protect\citeauthoryear{{Shaikhislamov}, {Fossati}, {Khodachenko}  \&
  et al}{{Shaikhislamov} et~al.}{2020}]{Shaikhislamov2020}
{Shaikhislamov} I.~F.,  {Fossati} L.,  {Khodachenko} M.~L.,   et al 2020, arXiv
  e-prints, \href {https://ui.adsabs.harvard.edu/abs/2020arXiv200606959S} {p.
  arXiv:2006.06959}

\bibitem[\protect\citeauthoryear{{Strubbe} \& {Chiang}}{{Strubbe} \&
  {Chiang}}{2006}]{2006ApJ...648..652S}
{Strubbe} L.~E.,  {Chiang} E.~I.,  2006, \mn@doi [\apj] {10.1086/505736}, \href
  {https://ui.adsabs.harvard.edu/abs/2006ApJ...648..652S} {648, 652}

\bibitem[\protect\citeauthoryear{{T{\'o}th} et~al.,}{{T{\'o}th}
  et~al.}{2005}]{Toth-swmf}
{T{\'o}th} G.,  et~al., 2005, \mn@doi [JGR (Space Physics)]
  {10.1029/2005JA011126}, \href
  {http://adsabs.harvard.edu/abs/2005JGRA..11012226T} {110, A12226}

\bibitem[\protect\citeauthoryear{{Vidal-Madjar}, {Lecavelier des Etangs},
  {D{\'e}sert}, {Ballester}, {Ferlet}, {H{\'e}brard}  \&
  {Mayor}}{{Vidal-Madjar} et~al.}{2003}]{VM2003}
{Vidal-Madjar} A.,  {Lecavelier des Etangs} A.,  {D{\'e}sert} J.-M.,
  {Ballester} G.,  {Ferlet} R.,  {H{\'e}brard} G.,   {Mayor} M.,  2003, \mn@doi
  [\nat] {10.1038/nature01448}, \href
  {https://ui.adsabs.harvard.edu/abs/2003Natur.422..143V} {422, 143}

\bibitem[\protect\citeauthoryear{{Vidotto} \& {Cleary}}{{Vidotto} \&
  {Cleary}}{2020}]{Vidotto2020}
{Vidotto} A.~A.,  {Cleary} A.,  2020, \mn@doi [\mnras] {10.1093/mnras/staa852},
  \href {https://ui.adsabs.harvard.edu/abs/2020MNRAS.494.2417V} {494, 2417}

\bibitem[\protect\citeauthoryear{{Vidotto} \& {Donati}}{{Vidotto} \&
  {Donati}}{2017}]{VidottoDonati2017}
{Vidotto} A.~A.,  {Donati} J.~F.,  2017, \mn@doi [\aap]
  {10.1051/0004-6361/201629700}, \href
  {https://ui.adsabs.harvard.edu/abs/2017A&A...602A..39V} {602, A39}

\bibitem[\protect\citeauthoryear{{Vidotto}, {Fares}, {Jardine}, {Moutou}  \&
  {Donati}}{{Vidotto} et~al.}{2015}]{Vidotto2015}
{Vidotto} A.~A.,  {Fares} R.,  {Jardine} M.,  {Moutou} C.,   {Donati} J.~F.,
  2015, \mn@doi [\mnras] {10.1093/mnras/stv618}, \href
  {https://ui.adsabs.harvard.edu/abs/2015MNRAS.449.4117V} {449, 4117}

\makeatother
\end{thebibliography}

\end{document}